\documentclass{llncs}

\usepackage{url}
\usepackage{pifont}
\usepackage{graphicx}

\usepackage{tikz}
\newcommand\TikCircle[1][1.6]{\tikz[baseline=-#1]{\draw[thick](0,0)circle[radius=#1mm];}}

\usepackage{oz,stmaryrd}
\newcommand{\st}{\,\cdot\,}

\newenvironment{reason}{\begin{tabbing}\hspace{2em}\=\kill}{\end{tabbing}\vspace{-2.5ex}}

\newcommand{\br}{\begin{reason}}
\newcommand{\er}{\end{reason}}

\newcommand{\of}{:\!}

\newcommand{\tbox}{{\thicklines\framebox(8,8){}}}
\newcommand{\tnext}{\raisebox{1mm}{\thicklines\TikCircle}}

\begin{document} 

\title{NFT formalised}
\author{Martha N.\,Kamkuemah\thanks{Corresponding author: {\tt martha@aims.ac.za}} 
\and J.\,W.\,Sanders}
\institute{AIMS South Africa, Cape Town, South Africa\\[2ex] \today} 

\maketitle

\begin{abstract}

Non-fungible tokens, NFT, have been used to record 
ownership of real estate, art, digital assets, and more recently to 
serve legal notice. They provide an important and accessible 
non-financial use of cryptocurrency's blockchain but are peculiar 
because ownership by NFT confers no rights over the asset. 
This work shows that it is possible to specify that peculiar 
property by combining functional and epistemic properties. 
Suitability of the specification is evaluated by proof that the
blockchain implementation conforms to it, and by its use in an 
analysis of serving legal notice.

\end{abstract}

\section{Introduction}
\label{sec.intro}

We begin by motivating the peculiar property that ownership by NFT 
confers no right over the asset. 
The combined evolutions of affluence and education 
has resulted in a trade-off between 
ownership of a unique good and its democritization. 
Recall, for example, the evolving ownership of reading 
matter. 

The oldest known documents, from the $35^{\rm th}$ 
century BC (see \cite{wiki}: List of oldest documents), 
were mostly religious or historical and for use by those 
in power. 
The Egyptian \textit{Books of the Dead} were religious
documents written for the pharaoh or queen by royal 
scribes; the Chinese \textit{Sh\^{u}} was an historical  
and religious collection compiled for the ruler. 
Contemporary populations were illiterate and poor, so 
that their owning such unique documents simply
did not arise. 

In the Middle Ages illiteracy was still widespread in 
Europe. Manuscripts were handwritten and often 
sumptuously illuminated, like the religious \textit{Books 
of Hours}. Each was unique and extremely valuable, 
and libraries were limited to religious orders or the very 
rich. Such manuscripts lay beyond the knowledge or 
interest of most of the population. 
Ownership was severely limited and documents unique.

In Europe, increased education and the printing press 
(Gutenberg, $\sim$1450) led ultimately to the era of 
newspapers (Strasbourg, 1605) and then in the last 
century to paperback books (Lane, Penguin Books, 1935) 
for literature, making reading matter affordable to virtually 
all, but at the expense of anything like the uniqueness
of documents. The 
market in first editions, and the continued popularity of 
author-signed copies, indicate the pleasure people 
derive from owning something distinctive and, where 
possible, unique 
(compare with the popularity of personalised car registration 
plates). 

Digital books began in the 1960s and 70s and in the
early 2020s are typified by the commercially successful 
Amazon \textit{Kindle} service \cite{kindle}. A digital file
is in principle able to be copied without limit. So 
ownership of  anything digital is very far from owning
something distinctive or unique.
Documents have evolved from being unique but beyond
ownership to being far from unique but owned. 

That evolution, and digital assets in particular, 
have required legal rights to keep pace. The 
\textit{public domain} is composed of all creative 
assets to which no exclusive intellectual property 
rights apply.
\textit{Copyright} gives its owner 
exclusive right over a creative asset to copy, 
distribute, adapt, display, and perform it.
To cope with copyable assets, \textit{Copyleft} 
grants certain freedom over copies 
of assets with the requirement that the same 
rights be preserved in derivatives to use for 
any purpose, to modify, copy, share, and 
redistribute it. (Condensed from \cite{wiki}: 
Public domain; Copyright; Copyleft.)

Since the advent of cryptocurrencies in 2008, and 
their distributed ledger the {blockchain}, the concept 
of \textit{non-fungible token}, or NFT, has offered a 
return to a kind of unique ownership though without 
any of those rights. The token itself is digital, 
appearing on a blockchain. But the asset to which it 
applies may be traditional, like real estate and 
sculpture, or digital, like computer art. Ownership, 
without rights and of something which can be 
copied without limit, is an interesting and subtle 
concept. 

NFT have captured popular imagination by 
allowing ownership of 
Cryptokitties \cite{kitties}; 
membership of the Bored Ape Yacht Club \cite{bayc}; 
ownership of digital art \cite{nftart}; 
or of the first tweet on Twitter \cite{twitter}.
They also facilitate a history of ownership which is
important for assets like diamonds \cite{diamonds} 
where an origin in conflict is to be avoided.
More recently in the United States and Britain, 
delivery of an NFT to an electronic wallet has become 
an acceptable form of serving legal notice \cite{law}.

Multifarious future uses of NFT are likely. So it 
seems sensible to have a specification of NFT to
which its blockchain implementation is shown to
conform. 
That is the purpose of this paper, which is arranged 
as follows. Ownership is studied in Section 
\ref{sec.owns} and used to specify NFT, by predicate 
$\cal N$, in Section \ref{sec.nft}. Account is taken of 
its epistemic properties phrased in terms of `public 
certifiability,' predicate $PC$. In Section \ref{sec.impl} 
the blockchain implementation is formalised and 
shown to satisfy the specification $\cal N$. 
Finally in Section \ref{sec.law} the NFT formalism, 
and particularly its public certifiability, are 
considered with the case study of serving legal 
notice.

As with many recent developments in Information 
Technology, traditional references are not always 
appropriate. For example the elegant and hugely 
influential whitepaper \cite{naka} by Satoshi 
Nakamoto which began Bitcoin, the first 
cryptocurrency (of which there are now over 
20,000) is not a refereed publication. References 
in this paper include both online documents and 
sites.

Background on NFT is covered by 
\cite{verge,investopedia} and
\cite{wiki}: Non-Fungible Tokens.

\section{Ownership}
\label{sec.owns}

Suppose \textit{Agent} denotes the type 
containing all owners and potential owners,  
assumed to be individuals; group ownership
is considered at the end of this section. 
Suppose \textit{Asset} denotes the type of all 
things regarded as assets, physical or digital, 
existing or future, and 
$Asset_{\scriptscriptstyle \exists}$ denotes 
the subset of those assets currently in 
existence. 

Time is assumed to have type $\mathbb{T}$ 
isomorphic to the natural numbers: there is 
an initial time, and the difference between
consecutive times is constant.

\begin{definition}
\label{def.owns}
{\em (Ownership)}\quad
The Boolean-valued function
\begin{eqnarray*}
Owns : Agent \times Asset \times \mathbb{T}\ \fun\ \bool
\end{eqnarray*}
is interpreted: $Owns(a,\alpha ,t)$ holds iff 
agent $a$ owns asset $\alpha$ at time $t$.
\end{definition} 

At any time not every agent need own an asset
and not every asset $\alpha \of Asset$ need be 
owned; but for each agent-asset pair, ownership 
is well defined. 

The set $Asset_{\scriptscriptstyle \exists}$ of
existing assets is temporal, so 
$Asset_{\scriptscriptstyle \exists}$ has an implicit 
time variable. Furthermore often the time variable 
of $Owns$ will be left implicit, and temporal 
operators used to express its temporal invariants, 
with\ \,$\tbox$\,\ for `now and in future', and\ 
\,$\tnext$\,\ for `next time' (well defined by our
assumption on $\mathbb{T}$). The usual 
laws of linear temporal logic hold \cite{logic}.

Each asset is \textit{minted} at some time, the 
point at which its default original owner identifies 
it by an ownership transaction on the blockchain.
It may subsequently be put up for sale on one of the 
various sites, depending on the nature of the 
asset. 

Subsequently the asset may change ownership many
times, involving standard transactions on the blockchain. 
Ownership at any time is established by searching 
back through the blockchain, resulting in the list of
owners (see Display (\ref{eq.search})), just as for 
validation of a financial transaction.

Elimination of an asset is not considered here,
though it is straightforward. 

Group ownership may be covered by replacing 
\textit{Agent} by
the set of all nonempty sets of agents. Then 
ownership by a set $S$ of agents may be 
thought of as ownership by the parallel 
composition of the members of $S$. The 
laws of ownership hold with 
individual owners replaced by nonempty 
sets of owners, as do: Definition \ref{def.nft} 
of NFT; the extensions of the operations 
$Mint$ and $TxO$; and the results.
(In fact the case of empty $S$ may be used 
as owner of a non-existing asset, to obviate
the need for $Asset_{\exists}$.)

\subsection{Laws}
\label{sec.laws}

Properties of \textit{Owns} are as follows. 
Each is stated by abstracting the time variable, 
and quantifying it instead with\ \,$\tbox$\, in 
order to focus on the property which remains 
invariant with time. The first three laws are 
fundamental.
\begin{enumerate}
\item
Each existing asset has an owner at any time:
\begin{eqnarray}
\label{eq.atleastone}
&& \tbox\ (\forall \alpha \of Asset_{\scriptscriptstyle \exists} \st \exists a \of Agent \st 
	Owns(a,\alpha ))\,. 
\end{eqnarray}

\item
At any time, each existing asset has at most one (hence exactly one) owner:
\begin{eqnarray}
\label{eq.atmostone}
&& \tbox\ (\forall \alpha \of Asset_{\scriptscriptstyle \exists} \st \forall a,a' \of Agent \st 
\left(\begin{array}{l}
Owns(a,\alpha )\ \wedge \\
Owns(a',\alpha )
\end{array}\right)
 \implies\ a = a' )\,.
\end{eqnarray}
An agent may own many assets, or none.

\item
A non-existing asset does not have an owner, since 
until it comes into existence it is not assumed to have
an identity:
\begin{eqnarray}
\label{eq.none}
&& \tbox\ (\forall \alpha \of Asset \setminus Asset_{\scriptscriptstyle \exists} \st 
	\neg \exists a \of Agent \st Owns(a,\alpha ))\,. 
\end{eqnarray}

\item
$Asset_{\scriptscriptstyle \exists}$ increases with time, since elimination of
assets is not considered:
\begin{eqnarray}
\label{eq.inc}
&& \tbox\ (Asset_{\scriptscriptstyle \exists}\, \subseteq\, \tnext\, Asset_{\scriptscriptstyle \exists} )\,. 
\end{eqnarray}

\item
Consequently, the size of $Owns$ increases with time (even
though that need not be true of the assets owned by any 
individual):
\begin{eqnarray}
\label{eq.incc}
&& \tbox\ (\# Owns\ \leq\, \tnext\, \# Owns)\,. 
\end{eqnarray}

\item
For any asset, the list of its past owners grows
(as a result, the commission earnt for its 
originator grows). In terms of one list being a
prefix of another:
\begin{eqnarray}
\label{eq.inccc}
&& \tbox\ (\,[a \of Agent \mid Owns(a,\alpha ) ]\,\ {\bf prefixes}\,\ \tnext\, [a \of Agent \mid Owns(a,\alpha ) ]\, )\,. 
\end{eqnarray}
\end{enumerate}

%
%
%

\section{NFT specified}
\label{sec.nft}

An NFT is a token on a blockchain. What properties 
should it have?

Firstly the token represents ownership of some asset
so at any time it must satisfy properties 
(\ref{eq.atleastone}) to (\ref{eq.inccc}), and in particular
the fundamental (\ref{eq.atleastone}) to (\ref{eq.none}). 
But furthermore, that ownership must be `publicly certified.' 

\textit{Certified} is interpreted to mean that all are aware 
of it. Without that, differences of awareness may occur; 
certifiability imposes uniformity of awareness.
\textit{Publicly} certified means that all are aware 
that others are aware of what is certified. Without that, 
ignorance may be feigned; being public ensures a kind 
of authenticity. (Common knowledge, which public 
certifiability approximates to depth 2, is not 
freshly achievable in any realistic distributed system.)

For example a decision by a governing body is abided by
not because the body is fortified by might or the law, 
but because of prior agreement and the fact that the 
board and its decision are publicly certified. 

Not every certification is publicly so. I have various friends
some of whom know each other and some of whom know
no others. If I email my website  to my friends, one-by-one, 
then each friend knows it. But none knows that the others 
know it. So amongst my friends my website is certified but
not publicly certified.

To express public certifiability, recall that the epistemic 
temporal formula $K_x\, \phi$ means that agent $x$ 
knows predicate $\phi$ (see \cite{logic}). For variable 
$v \of \mathbb{V}$, \,$K_x\, v$\,\ is shorthand for 
$\exists w : \mathbb{V} \st  K_x\, ( v = w )$. 

In epistemic logic only truths can be known: if\,\
$\vdash K_x\, \phi$\ \,then\,\ $\vdash \phi$. 

In terms of epistemic logic, 
\begin{definition}
\label{def.pc}
{\em (Public certifiability)}\quad
Fact $\phi$ is {\em publicly certified} amongst a set 
$A$ of agents: 
\begin{eqnarray*}
PC(A,\phi) &\ :=\ & \forall x,y \of A \st K_x K_y \,\phi \,. 
\end{eqnarray*}
\end{definition}

In particular, by the laws of epistemic logic, both
\((\forall y \of A \st K_y \,\phi)\) and \(\phi\) hold.

The final ingredient required is that of a token function 
to represent the relation $Owns(a, \alpha , t)$. Suppose 
$\tau$ is an injection
\begin{eqnarray*}
\tau : \{ (a,\alpha ,t) \of Agent \times Asset \times \mathbb{T} \mid 
Owns(a,\alpha ,t) \} \inj \bool^*
\end{eqnarray*} 
to bitstrings. The token function and its inverse are 
common knowledge and are both quick to compute. 
The function $\tau$ incorporates the identities of 
agent $a$ and asset $\alpha$, and assumes some 
global convention for time $t$, thus saving those from 
having to be considered in greater detail here. 

With that preparation:
\begin{definition}
\label{def.nft}
{\em (Specification \(\cal N\))}\quad
A non-fungible token, or NFT, is a publicly 
certified statement that agent $a$ owns 
asset $\alpha$ at time $t$:
\begin{eqnarray*}
PC(Agent, \tau(Owns(a,\alpha ,t))) \,,
\end{eqnarray*}
where the $Owns$ relation satisfies (\ref{eq.atleastone}) 
to (\ref{eq.none}). That is referred to as property \(\cal N\).
\end{definition}

The success of using a token function is due to:
\begin{eqnarray*}
K_x \tau (Owns(a, \alpha ,t)) &\ \iff\ & K_x Owns(a, \alpha ,t)\,. 
\end{eqnarray*}
Furthermore by the laws of epistemic logic, Property 
\(\cal N\) implies that each agent knows the value
$Owns(a,\alpha ,t)$. 

Although ownership is tokenised, ownership of a token 
for ownership is not considered, as that would entail an
infinite regress.

It is worth repeating: ownership by NFT does not 
confer copyright, so is an unusual kind of ownership. 
(What would be gained by buying the NFT for the pdf 
of this paper?)

\section{Blockchain implementation}
\label{sec.impl}

NFT has been specified in Definition \ref{def.nft} 
by property $\cal N$, independent 
of implementation. 

An implementation which is centralised, and hence 
of little practical interest, would be provided by a 
publicly accessible bulletin board maintaining an 
up-to-date list of which agent owns what asset. 
Being centralised and openly accessible its 
contents are publicly certified. But it would 
suffer the usual disadvantages of centralisation 
(including inefficiency of access due to 
bottlenecks, fragility, and susceptibility to 
corruption). 

A distributed solution is of course provided by  
blockchain. Different NFT platforms offer different
functionalities \cite{markets}. Assume a blockchain, 
$bc$, which supports the standard financial 
transactions and is sustained by a network 
$Net$ of nodes each running the blockchain 
software and holding a copy of $bc$ itself. 

Agents interact with $Net$ using e-wallets in the
standard manner. For simplicity it is assumed
that all $Net$ nodes (and wallets) are honest 
and execute the same $bc$ software; each 
knows that and moreover knows that the others 
know it. In other words: 
\begin{eqnarray}
\label{eq.pcbc}
PC(Net,\mbox{`all nodes are honest and run $bc$ software'})\,.
\end{eqnarray} 
So $Net$ nodes know they have the same version 
of $bc$ at any time.

Public certifiability is closed under subset of agents: 
if $PC(A,\phi)$ and $B \subseteq A$ then $PC(B, \phi)$.
However it is not closed under union. So in extending 
$PC(Net,\phi)$ to $PC(Agent,\phi)$ in Theorem 
\ref{thm.corr} below the following result will be helpful.

\begin{lemma} {\em (Extending PC)}\quad
\label{lem.trust}
Let $A$ be a set of agents and agent $x \notin A$.
If $PC(A,\phi)$ and all agents are honest 
(tell only the truth), then $PC(A \cup \{x \},\phi)$ 
provided for some $v_x \in A$, 
\begin{eqnarray*}
v_x \fun x &:& PC(A,\phi) \\
v_x \fun A &:& K_x \phi \,.
\end{eqnarray*}
\end{lemma}

\noindent{\bf Proof.}\quad
Establishing $PC(A \cup \{x\}, \phi)$ requires, by
Definition \ref{def.pc},
\begin{eqnarray*}
\forall v,w : A \cup \{x\} \st K_v K_w\, \phi \,.
\end{eqnarray*}
The cases $v \neq x$ and $w \neq x$ are covered 
by assumption. 

If $v = x$ then from the first assumed
communication, $\forall w : A \st K_x K_w\, \phi $.

If $w = x$ then from the second assumed
communication and liveness, 
$\forall v : A \st K_v K_x \,\phi$.
\hfill$\Box$\\

For simplicity it is also assumed of $bc$ 
that each block consists of a single 
transaction, either standard or an 
ownership token with its associated 
standard transactions, and that 
appending a block takes one time unit.
In practice transaction fees are also 
included to ensure quick inclusion in a 
block. Those are neglected here, and 
relegated to the protocol between an 
e-wallet and the memory pool of its 
local $Net$ node. 

However $Net$ communication is
publicly certified and can be thought of
as a distributed clock which is 
self-regulated to tick every ten minutes 
on average.

An NFT is implemented by a token on $bc$, 
stating ownership of an asset by an agent at 
the time of the token's inclusion in $bc$. To 
manage NFT, what properties must $bc$ 
maintain?

\subsection{Minting}
\label{sec.mint}

Currently there seems to be no restriction
on the asset or minter of an NFT.

Operation $Mint(orig,\alpha ,t\,{+}\,1)$, by which 
agent $orig$ establishes original ownership of 
asset $\alpha$ at time $t\,{+}\,1$, inputs 
$orig \of Agent$, $\alpha \of Asset$ and 
$t \of \mathbb{T}$. If $\alpha$ is not already 
owned at $t$, it appends to $bc$ a block with 
token $\tau(Owns (orig,\alpha ,t\,{+}\, 1))$
representing original ownership of $\alpha$ by $a$. 
But if $\alpha$ is owned at $t$  then an error
message is returned and $bc$ is left unchanged. 

The precondition for $Mint$ to update $bc$ is:
\begin{eqnarray*}
\neg \exists a \of Agent \st \exists u \of \mathbb{T} \st u \leq t\ \land\ Owns(a,\alpha ,u) 
\end{eqnarray*}
in which case $bc$ is updated:
\begin{eqnarray*}
Append\,(\, \tau(Owns (orig,\alpha ,t\,{+}\, 1)),\, bc\, ) \,,
\end{eqnarray*}
where  the operation $Append\,(b,bc)$ appends block 
$b$ to chain $bc$.
If the precondition fails then $bc$ is unchanged.

\subsection{Standard transaction, $Tx$}
\label{sec.tx}

A standard transaction $Tx\,(b,s,c,t)$ on $bc$ involves buyer 
$b \of Agent$, seller $s \of Agent$, cost $c \of \real$ and time 
$t \of \mathbb{T}$. The asset changing hands in a standard 
transaction is of no concern to the $bc$ ledger, which is 
occupied entirely with the financial validity of the transaction. 
That requires a user's sales to exceed its purchases by at 
least the cost $c$ of the current transaction, a fact which is 
determined by a scan through $bc$: 
\begin{eqnarray} 
\label{eq.balance}
\sum \{ d \mid \exists b \!\st\! Tx\,(b,u,d)\ {\bf in}\ bc \} \, -\,
\sum \{ d \mid \exists s \!\st\! Tx\,(u,s,d)\ {\bf in}\ bc  \}\ \geq\, c \,.
\end{eqnarray}
In that case (only), the transaction is appended to $bc$. 

Deposit, and reward for adding a block to $bc$, are considered 
to be sales; and withdrawal is considered to be a purchase.

\subsection{Change of ownership transaction, $TxO$}
\label{sec.otx}

An ownership transaction $TxO$ extends a standard transaction
$Tx$ by making the asset, $\alpha$, explicit. It inputs the current
owner $old$ and buyer $new$, the asset $\alpha \of Asset$, 
cost $c \of \real$ and time $t \of \mathbb{T}$. Its precondition is 
that $new$ can afford the cost $c$ (in standard manner) and 
$old$ does indeed own $\alpha$ at that time: 
\begin{eqnarray*}
Owns (old,\alpha ,t)\ \land\ balance(new) \geq c
\end{eqnarray*}
where $balance(new)$ equals the left-hand side of
Inequality (\ref{eq.balance}).

In that case a standard transaction is invoked for the sale. 
It is concatenated to the ownership token and so included 
in the block which is appended to $bc$:  
\begin{eqnarray*}
Append\,{\bf (}\, \tau(Owns (new,\alpha ,t\,{+}\, 1))\, {\bf concat} Tx(new , old , c , t\,{+}\, 1) ,\, bc\, {\bf )}\,.
\end{eqnarray*}
Again, $TxO$ takes one time unit to append the block 
containing that catenation.
 
 A block in $bc$ contains either a standard transaction
 or a $Mint$ or $TxO$ transaction (which includes a
 standard transaction). To check the precondition of a
 standard transaction all of $bc$ must be searched.
 But for ownership only those blocks involving tokens
 need be checked, information which may be included 
 in the block header.
 
 Royalties are considered in Section \ref{sec.royal}.

\subsection{Correctness}
\label{sec.corr}

$Asset_{\scriptscriptstyle \exists}$ is evidently represented 
in the blockchain implementation as follows. At any time 
$t \of \mathbb{T}$: 
\begin{eqnarray}
\label{eq.Ae}
Asset_{\scriptscriptstyle \exists} & = & \{ \alpha \of Asset \mid \exists u \of \mathbb{T} 
	\st u \leq t\ \land\  \exists a \of Agent \st Mint(a,\alpha ,u) \} \,.
\end{eqnarray}

As strong as the epistemic condition of public 
certifiability might seem, it holds for the distributed 
implementation using blockchain:

\begin{theorem} {\em (Correctness)}\quad
\label{thm.corr}
The blockchain implementation of NFT, assuming
$Net$ nodes are honest and incorporating
$Mint$, $TxO$ and the communications of Lemma
\ref{lem.trust}, satisfies property $\cal N$ of 
Definition \ref{def.nft}. 
\end{theorem}

\noindent{\bf Proof.}\quad
For any asset $\alpha \of Asset$, a search of $bc$ by 
any node in $Net$ returns the same list of owners, 
ordered chronologically in the same way as $bc$,
\begin{eqnarray}
\label{eq.search}
as(\alpha) & := & [ \,a \of Agent \mid \tau[Owns(a,\alpha)]\ {\bf in}\ bc\, ] \,.
\end{eqnarray}
Thus at any height of $bc$, all nodes have the same 
value for $Owns(a,\alpha)$.

The proof establishes $bc \models {\cal N}$ by 
induction on $bc$. First consider just properties 
(\ref{eq.atleastone}) to (\ref{eq.none}) of $Owns$.

For the base case, $bc = [\ ]$, $Owns = \emptyset$ 
which vacuously satisfies the three properties. 

For the step case suppose that 
$bc = Append(b,bc_0)$ with block $b$ containing 
a token transaction. Either $\# bc_0 = 0$ or 
$\# bc_0 > 0$. 
In the first case, $Mint$ must have occurred 
since with $bc = [\ ]$ it is the only token action
whose precondition holds. Thus $b$ contains 
$\tau[Owns(orig,\alpha)]$, and $Owns$ satisfies
(\ref{eq.atleastone}) to (\ref{eq.none}). 
In the second case, $bc = Append(b,bc_0)$ 
where $\# bc_0 > 0$ satisfies the three 
properties. The last token action to occur in 
$bc_0$ was $Mint$ if $\# bc_0 = 1$ and 
otherwise $TxO$. 
In either case the three properties hold by 
the invariant maintained by definition of 
both operators. 

The proof of public certifiability is the
same for both the base and step cases
of that induction, so is given once. 
From (\ref{eq.pcbc}), Equations (\ref{eq.Ae}) 
and (\ref{eq.search}) imply that the value 
of $Owns(a,\alpha)$ is publicly certified
throughout $Net$: $PC(Net,Owns)$.
That is extended from $Net$ to 
$Agent$ as follows.
For each $x : Agent$ its e-wallet
connects to some $v_x \in Net$. Since
the conditions of Lemma \ref{lem.trust}
hold, $PC(Net \cup \{x\}, Owns)$. 
Iterating, $PC(Agent, Owns)$, so 
$bc \models {\cal N}$ as required. 
\hfill$\QED$

\subsection{Royalties transaction, $TxOr$}
\label{sec.royal}

Some NFT platforms offer the originator a royalty with 
each change in ownership. 
If $c$ is the total cost of an ownership change then the 
royalty, written $\% c$, is paid by $new$ to $orig$ with 
a standard transaction, just like its payment of the 
deficiency $(c{-}\% c)$ to $old$.

The transaction for change of ownership, but incorporating
royalties, is written $TxOr$. The originator of 
$\alpha \of Asset$ is not an input since it is identified 
by any node as the first element of $as(\alpha)$ in 
(\ref{eq.search}).

$TxOr$ has the same precondition as $TxO$ (Section 
\ref{sec.otx}) and the same update of $bc$ with an 
ownership token. However
its successful occurrence invokes an additional  
standard transaction for the royalty, concatenated 
with the standard transaction of $TxO$:
\begin{eqnarray*}
Tx\,(new,old,c{-}\% c,t\,{+}\, 1)\, {\bf concat}\, Tx\,(new,orig,\% c,t\,{+}\, 1)\,.
\end{eqnarray*}
In fact those standard transactions may occur in either order.
The block then contains the concatenation with the token.

The correctness of the modified system follows
as for the Correctness Theorem.

\begin{theorem} {\em (Corollary)}\quad
The inclusion of royalties in ownership transactions,
replacing $TxO$ by $TxOr$, conforms to ${\cal N}$.
\end{theorem}

\section{Case study: serving legal notice}
\label{sec.law}

The serving of a legal notice is a formal action 
requiring stringent conditions about which there 
seems to be approximate international agreement 
(see \cite{law2} and \cite{wiki}: Service of a 
Process). There are several possible standard 
methods, including service by:

\noindent ($\alpha$)\quad 
an officer of the court directly to the recipient; 

\noindent ($\beta$)\quad 
registered post to the recipient;

\noindent ($\gamma$)\quad 
email to the recipient's email address;

\noindent ($\delta$)\quad 
publication in a newspaper's public notices;
\vspace{.8ex}

Typically, $\alpha$ is used if possible. 
When that is infeasible, $\beta$ 
is acceptable but $\gamma$ is not. 
If the court does not have an address 
for the recipient, $\delta$ may be used. 

Recently \cite{law}, courts in both the 
US and UK have allowed service by: 
\vspace{.8ex}

\noindent ($\varepsilon$)\quad 
NFT to an e-wallet. 
\vspace{.8ex}

Method $\varepsilon$ provides a quite 
different use of NFT. It also diverges 
from the acceptable methods $\alpha$, 
$\beta$ and $\delta$ of serving notice
and resembles closely the unacceptable 
method $\gamma$. So it is included 
here as a case study; see Theorem
\ref{thm.sln}.

What role do functional and epistemic 
properties play in the serving of notice?
The standard methods are included and
considered first, for comparison with 
$\varepsilon$.

Consider a population $Pop$ with court $cc$, 
and a notice $nn$ to be served to recipient 
$rr : Pop$. Let
\vspace{-1.5ex}
\begin{eqnarray*}
\phi_0 &\ :=\ & cc \mbox{ has authority} \\
\phi_1 &\ :=\ & (\mbox{notice} = nn) \\
\phi_2 &\ :=\ & \mbox{the serving officer has the authority of } cc \\
\phi_3 &\ :=\ & rr \mbox{ has been served with } nn.
\end{eqnarray*}
A condition required by a successful court 
system can be expressed as $\phi_0$ being 
publicly certified amongst $Pop$, namely 
$PC(Pop, \phi_0 )$.

The following four derived properties are 
used to answer the previous question. 

The recipient $rr$ knows the authority of
the method of being served notice:
\begin{eqnarray}
&& K_{rr}\, \phi_2 \label{eq.(a)} \,. 
\end{eqnarray} 
The court knows that $rr$ has been served 
with $nn$:
\begin{eqnarray}
&& K_{cc}\, \phi_3 \label{eq.(b)} 
\end{eqnarray} 
and recipient $rr$ knows that fact:
\begin{eqnarray}
&& K_{rr}\, K_{cc} \,\phi_3 \label{eq.(c)} \,.
\end{eqnarray}
Furthermore the court knows that $nn$ 
has been served privately. The only
way for others to know about $nn$ is by
$rr$ telling them. So ruling out $rr$'s 
private communications after receiving 
$nn$,
\begin{eqnarray}
&& K_{cc} ( (K_x \,\phi_3)\ \implies\ x \,{=}\, rr) \label{eq.(d)} \,.
\end{eqnarray} 

The standard methods satisfy these properties: 

\begin{lemma} {\em (Standard methods)}\quad
\label{lem.serve}
Service of notice by method $\alpha$ or by 
method $\beta$ achieve Properties (\ref{eq.(a)}) 
to (\ref{eq.(d)}). 
Method $\gamma$, by email, satisfies just
Properties (\ref{eq.(a)}) and (\ref{eq.(d)}).
Method $\delta$, by newspaper publication,
achieves none of (\ref{eq.(a)}) to (\ref{eq.(d)}).
\end{lemma}

\begin{center}
\begin{tabular}{|c||c|c|c|c|} \hline
Method 		&\ (\ref{eq.(a)})\ \ &\ (\ref{eq.(b)})\ \ 		&\ (\ref{eq.(c)})\ \ &\ (\ref{eq.(d)})\ \ \\ \hline\hline
$\alpha$ 		& \checkmark	    & \checkmark      		& \checkmark  	   & \checkmark  \\
$\beta$ 		& \checkmark   	    & \checkmark      		& \checkmark       & \checkmark  \\
$\gamma$ 	& \checkmark 	    & \ding{55}          		& \ding{55}           & \checkmark  \\
$\delta$ 		& \ding{55}   	    & \ding{55}          		& \ding{55}           & \ding{55}      \\ \hline
$\varepsilon$ 	& \checkmark   	   & \checkmark 		& \checkmark       & \checkmark  \\ \hline
\end{tabular}
\end{center}

\noindent{\bf Proof \&\,discussion.}\quad 
Consider the four methods individually.
\vspace{.8ex}

($\alpha$) In the method of delivery by
officer of the court directly to $rr$, the 
serving officer shows publicly 
certified documentation indicating 
$\phi_2$. In the distributed setting such 
documentation consists of the public key 
of the officer (his or her ID) signed by the 
private key of $cc$. Either way, 
(\ref{eq.(a)}) is achieved.
%
The serving of $nn$ directly to $rr$ means  
that $cc$ knows the notice has been served,
(\ref{eq.(b)}), and $rr$ knows that $cc$ 
knows that fact (ruling out any pretence by
$rr$ at ignorance), (\ref{eq.(c)}); finally it 
maintains privacy in the sense of (\ref{eq.(d)}).
\vspace{.8ex}

($\beta$) The method of registered post 
satisfies (\ref{eq.(a)}) because the system of 
registered post is itself publicly certified and 
the sender, $cc$, is revealed on delivery. 
(Formalisation of the system of registered 
post and verification that it is publicly
certified requires no new concepts and is 
omitted.)
Properties (\ref{eq.(b)}) and (\ref{eq.(c)}) 
hold because $rr$ has to sign for 
receipt, a fact which is on record. 
(\ref{eq.(d)}) holds because the delivery is 
made to only $rr$. 
\vspace{.8ex}

($\gamma$) For the method of service
by email, presumably $nn$ contains 
reference to $rr$ and is encrypted with 
both $cc$'s private key and $rr$'s public 
key. So $rr$ knows the message originated 
with $cc$ and and cannot have been read 
and relayed (as in a man-in-the-middle 
attack). So Properties (\ref{eq.(a)}) and 
(\ref{eq.(d)}) hold.
Property (\ref{eq.(b)}) fails because of
communication delays and failures.
Since all users know that, the
second-order (\ref{eq.(c)}) also fails.
\vspace{.8ex}

($\delta$) The method of newspaper
publication meets none of (\ref{eq.(a)}) to 
(\ref{eq.(d)}). It may be thought of as a 
way of being seen publicly to be attempting 
to serve notice in the absence of any better
method. 
\hspace*{\fill}$\QED$\\

The properties of $\varepsilon$ follow 
using the reasoning established in Lemma 
\ref{lem.serve}.

\begin{theorem} {\em (Service by NFT)}\quad
\label{thm.sln}
Method $\varepsilon$, by NFT to an 
e-wallet, achieves Properties (\ref{eq.(a)}) 
to (\ref{eq.(d)}), with qualification against 
(\ref{eq.(b)}) and (\ref{eq.(c)}).
\end{theorem}

\noindent{\bf Proof \&\,discussion.}\quad  
Consider service by NFT---representing 
ownership by $cc$ of asset $nn$---to an 
e-wallet $rr$. Now (\ref{eq.(a)}) holds 
due to the public certifiability of $cc$,
$PC(Pop, \phi_0 )$, and properties 
$\cal N$.

Condition (\ref{eq.(b)}), that $cc$ knows 
that $rr$ has been served with $nn$, 
might be thought to fail for the same 
reasons it does with $\gamma$: 
network communication is 
asynchronous and subject to
uncertified failure. However in the
case of $\varepsilon$, communication
is from e-wallet to e-waller \textit{via} the
blockchain $Net$. The reliability of such 
communications is publicly certified, as 
observed in Section \ref{sec.impl}.

A further qualification is necessary. An 
e-wallet, identified by a public key, is 
unique. But the identity of the owner(s) 
is in general unknown. In such cases 
it is the e-wallet itself which is being 
served 
(for instance in \textit{D'Aloia v (1) 
Persons Unknown and (2) Binance 
Holdings Limited and others}, \cite{law}). 
Since there is
no alternative when it is the anonymous 
owners of the e-wallet who are actually 
being served. With those qualifications, 
(\ref{eq.(b)}) holds.
(Aside: supposing the e-wallet `knows' 
of the service of notice, does any 
human?)

Condition (\ref{eq.(c)}), that $rr$ knows 
that $cc$ knows, holds as above 
because communication on $Net$ is 
publicly certified. Condition 
(\ref{eq.(d)}), that privacy is preserved, 
holds as in $\gamma$ assuming all
those with access to the e-wallet $rr$ 
are being served.
\hfill$\QED$\\

Ownership documented by an NFT 
and its public certifiability, provides a 
convenient way to ensure authority. 
In that case study, it is the authority 
of the court or its representative.

\section{Conclusion}

{\bf In summary} this work has specified the 
idea of NFT by combining functional and 
epistemic properties: unique ownership of 
an asset and public certifiability $PC$ 
(common knowledge approximated to 
depth 2) of that ownership, $\cal N$. It 
has formalised the blockchain 
implementation and shown it to conform 
to specification $\cal N$, and shown that 
service of a legal notice can be reasoned 
about using those ideas.

{\bf In conclusion} the combination of
functional and epistemic properties seems
appropriate, and has the advantage of 
being supported by algebraic reasoning 
without the need for semantic arguments.

{\bf Further work} includes the wider
application of $PC$ as a specification
technique in community-based distributed 
systems. And investigation of what 
might be called SNFT, or \textit{smart 
non-fungible tokens}, representing 
ownership which changes under certain 
conditions (a smart contract). A simple 
example would be an asset whose 
ownership requires an annual payment 
(like a personalised car number plate).

\bibliography{nft}{}
\bibliographystyle{plain}

\end{document}